\documentclass[twocolumn]{aastex62}

\def\apjl{{ApJ}}	
\def\apjs{{ApJS}}	
\def\aap{{A\&A}}

\newcommand{\Kepler}{\emph{Kepler}\xspace}
\newcommand{\TESS}{\emph{TESS}\xspace}
\usepackage{xspace}
\newcommand{\rhocirc}{\rho_{\rm circ}}
\newcommand{\thisstar}{TOI-216\xspace}
\newcommand{\thisstarinn}{TOI-216b\xspace}
\newcommand{\thisstarout}{TOI-216c\xspace}
\newcommand{\inn}{b\xspace}
\newcommand{\out}{c\xspace}
\shorttitle{\thisstar}
\shortauthors{Dawson et al.}
\usepackage{multirow}

\begin{document}

\title{TOI-216b and TOI-216c: Two warm, large exoplanets in or slightly wide of the 2:1 orbital resonance}
\received{March 12, 2019}
\revised{April 25, 2019}
\correspondingauthor{Rebekah I. Dawson}
\email{rdawson@psu.edu}

\author[0000-0001-9677-1296]{Rebekah I. Dawson}
\affiliation{Department of Astronomy \& Astrophysics, Center for Exoplanets and Habitable Worlds, The Pennsylvania State University, University Park, PA 16802, USA}
\author[0000-0003-0918-7484]{Chelsea~ X.~Huang}
\affiliation{Department of Physics and Kavli Institute for Astrophysics and Space Research, Massachusetts Institute of Technology, Cambridge, MA 02139, USA}
\affiliation{Juan Carlos Torres Fellow}
\author{Jack J. Lissauer}
\affiliation{NASA Ames Research Center, Moffett Field, CA 94035}
\author[0000-0001-6588-9574]{Karen A.\ Collins}
\affiliation{Harvard-Smithsonian Center for Astrophysics, 60 Garden St., Cambridge, MA, 02138, USA}
\author[0000-0001-5401-8079]{Lizhou Sha}
\affiliation{Department of Physics and Kavli Institute for Astrophysics and Space Research, Massachusetts Institute of Technology, Cambridge, MA 02139, USA}
\author{James Armstrong}
\affiliation{University of Hawaii Institute for Astronomy, 
Pukalani, HI 96768}
\author[0000-0003-2239-0567]{Dennis M.\ Conti}
\affiliation{American Association of Variable Star Observers, 49 Bay State Road, Cambridge, MA 02138, USA}
\author[0000-0003-2781-3207]{Kevin I.\ Collins}
\affiliation{Department of Physics and Astronomy, Vanderbilt University, Nashville, TN 37235, USA}
\author{Phil Evans}
\affiliation{El Sauce Observatory, Coquimbo Province, Chile}
\author{Tianjun Gan}
\affiliation{Physics Department and Tsinghua Centre for Astrophysics, Tsinghua University, Beijing 100084, China}
\author[0000-0003-1728-0304]{Keith Horne}
\affiliation{SUPA Physics \& Astronomy, University of St~Andrews, North Haugh, St~Andrews, KY16~9SS, Scotland, UK}
\author{Michael Ireland} 
\affiliation{Research School of Astronomy and Astrophysics, Australian National University, Canberra, ACT 2611, Australia}
\author[0000-0001-9087-1245]{Felipe Murgas}
\affiliation{Instituto de Astrofísica de Canarias (IAC), E-38205 La Laguna, Tenerife, Spain}
\affiliation{Departamento de Astrofísica, Universidad de La Laguna (ULL), E-38206 La Laguna, Tenerife, Spain}
\author[0000-0002-9810-0506]{Gordon Myers} 
\affiliation{AAVSO, 5 Inverness Way, Hillsborough, CA 94010, USA}
\author{Howard M. Relles}
\affiliation{Harvard-Smithsonian Center for Astrophysics, 60 Garden St., Cambridge, MA, 02138, USA}
\author{Ramotholo Sefako}
\affiliation{South African Astronomical Observatory, PO Box 9, Observatory, 7935, South Africa}
\author[0000-0002-1836-3120]{Avi Shporer}
\affiliation{Department of Physics and Kavli Institute for Astrophysics and Space Research, Massachusetts Institute of Technology, Cambridge, MA 02139, USA}
\author[0000-0003-2163-1437]{Chris Stockdale}
\affiliation{Hazelwood Observatory, Australia}
\author[0000-0001-6023-4974]{Maru{\v s}a {\v Z}erjal}
\affiliation{Research School of Astronomy and Astrophysics, Australian National University, Canberra, ACT 2611, Australia}
\author{George Zhou}
\affiliation{Harvard-Smithsonian Center for Astrophysics, 60 Garden St., Cambridge, MA, 02138, USA}
\author{G. Ricker}
\affiliation{Department of Physics and Kavli Institute for Astrophysics and Space Research, Massachusetts Institute of Technology, Cambridge, MA 02139, USA}
\author[0000-0001-6763-6562]{R. Vanderspek}
\affiliation{Department of Physics and Kavli Institute for Astrophysics and Space Research, Massachusetts Institute of Technology, Cambridge, MA 02139, USA}
\author{David W. Latham}
\affiliation{Harvard-Smithsonian Center for Astrophysics, 60 Garden St., Cambridge, MA, 02138, USA}
\author[0000-0002-6892-6948]{S. Seager}
\affiliation{Department of Physics and Kavli Institute for Astrophysics and Space Research, Massachusetts Institute of Technology, Cambridge, MA 02139, USA}
\affiliation{Department of Earth, Atmospheric, and Planetary Sciences, Massachusetts Institute of Technology, Cambridge, MA 02139, USA}
\affiliation{Department of Aeronautics and Astronautics, Massachusetts Institute of Technology, Cambridge, MA 02139, USA}
\author[0000-0002-4265-047X]{J. Winn}
\affiliation{Department of Astrophysical Sciences, Princeton University, 4 Ivy Lane, Princeton, NJ 08540, USA}
\author{Jon M. Jenkins}
\affiliation{NASA Ames Research Center, Moffett Field, CA 94035}
\author[0000-0002-0514-5538]{L. G. Bouma}
\affiliation{Department of Astrophysical Sciences, Princeton University, 4 Ivy Lane, Princeton, NJ 08540, USA}
\author{Douglas A. Caldwell}
\affiliation{NASA Ames Research Center, Moffett Field, CA 94035}
\affiliation{SETI Institute, Mountain View, CA 94043, USA}
\author[0000-0002-6939-9211]{Tansu~Daylan}
\affiliation{Department of Physics and Kavli Institute for Astrophysics and Space Research, Massachusetts Institute of Technology, Cambridge, MA 02139, USA}
\affiliation{Kavli Fellow}
\author{John~P.~Doty}
\affiliation{Noqsi Aerospace Ltd., 2822 South Nova Road, Pine, CO 80470, USA}
\author{Scott Dynes}
\affiliation{Department of Physics and Kavli Institute for Astrophysics and Space Research, Massachusetts Institute of Technology, Cambridge, MA 02139, USA}
\author[0000-0002-9789-5474]{Gilbert A. Esquerdo} 
\affiliation{Harvard-Smithsonian Center for Astrophysics, 60 Garden St., Cambridge, MA, 02138, USA}
\author[0000-0003-4724-745X]{Mark Rose}
\affiliation{NASA Ames Research Center, Moffett Field, CA 94035}
\author[0000-0002-6148-7903]{Jeffrey C. Smith}
\affiliation{NASA Ames Research Center, Moffett Field, CA 94035}
\affiliation{SETI Institute, Mountain View, CA 94043, USA}
\author[0000-0003-1667-5427]{Liang~Yu}
\affiliation{Department of Physics and Kavli Institute for Astrophysics and Space Research, Massachusetts Institute of Technology, Cambridge, MA 02139, USA}

\begin{abstract}
Warm, large exoplanets with 10--100~day orbital periods pose a major challenge to our understanding of how planetary systems form and evolve. Although high eccentricity tidal migration has been invoked to explain their proximity to their host stars, a handful reside in or near orbital resonance with nearby planets, suggesting a gentler history of in situ formation or disk migration. Here we confirm and characterize a pair of warm, large exoplanets discovered by the \TESS Mission orbiting K-dwarf TOI-216. Our analysis includes additional transits and transit exclusion windows observed via ground-based follow-up. We find two families of solutions, one corresponding to a sub-Saturn{-mass planet} accompanied by a Neptune{-mass planet} and the other to a Jupiter in resonance with a sub-Saturn{-mass planet}. We prefer the second solution based on the orbital period ratio, the planet radii, the lower free eccentricities, and libration of the 2:1 resonant argument, but cannot rule out the first. The free eccentricities and mutual inclination are compatible with stirring by other, undetected planets in the system, particularly for the second solution.  We discuss prospects for better constraints on the planets' properties and orbits through follow-up, including transits observed from the ground.
\end{abstract}

\section{Introduction}

Warm large exoplanets, giant planets with 10--100~day orbital periods, pose a major challenge to our understanding of how planets form and evolve. Origins hypotheses developed and fine-tuned to account for the more readily discovered hot Jupiters (orbital periods $<10$~days) and the far more abundant warm sub-Neptunes find it challenging to account for warm, large exoplanets' occurrence rates, eccentricities, masses, and companion properties (e.g., \citealt{wu11,beau12,petr15,daws15b,huan16}; see Section~4.3 of \citealt{daws18} for a review). Although rarer than smaller planets and more distant giants, warm, large exoplanets are an outcome of physical processes that likely sculpt many planetary systems. 

\begin{figure}
\begin{center}
\includegraphics[width=3.5in]{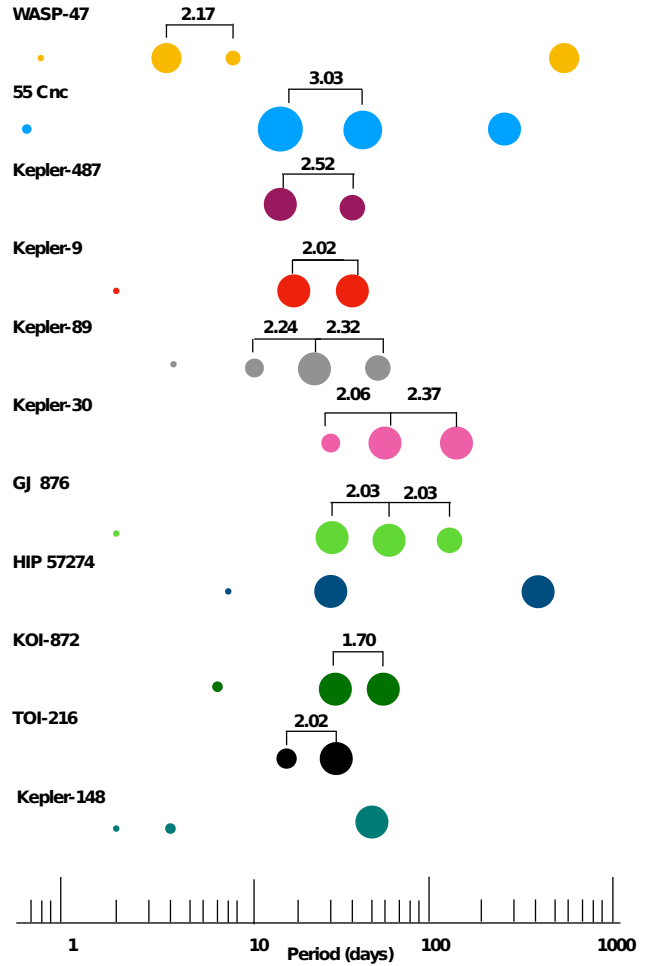}
\caption{
 \label{fig:arch} 
 All confirmed exoplanet systems with a warm, large exoplanet (mass greater than 0.25~$M_{\rm Jup}$ or radius greater than 8 Earth radii; orbital period less than 100~days) and one or more companions with a $<100$~day orbital period.  { (The WASP-47 system satisfies this criteria but contains a hot Jupiter.)}  Sizes shown are roughly proportional to planet size.
}
\end{center}
\end{figure}

Recently some have argued for two origins channels for warm, large exoplanets (e.g., \citealt{daws13,dong14,daws15,petr16}): high eccentricity tidal migration, and a second channel that may involve disk migration and/or in situ formation. Under the hypothesis of high eccentricity tidal migration, warm, large exoplanets are planets caught in the act of migration: they began further from the star, were disturbed onto highly elliptical orbits, and are tidal circularizing to short orbital periods. However, a key piece of evidence supporting the second channel is the handful of warm, large exoplanets with nearby planets, which are incompatible with high eccentricity migration { and are not on route to becoming hot Jupiters}.  Fig.~\ref{fig:arch} shows all confirmed systems with a warm, large exoplanet (mass greater than 0.25~$M_{\rm Jup}$ or radius greater than 8 Earth radii; period less than 100~days) and a companion with a $<100$~day orbital period. It is striking that most of these systems are in or near an orbital resonance, and almost all contain a known small planet on a $<10$~day orbital period, despite the low occurrence rate of such short period planets in general (e.g., \citealt{muld15}). They also happen to be some of the most iconic, well-studied exoplanet systems, probably because large and/or massive planets with short orbital periods are most amenable to transit and radial velocity characterization. Discovering and characterizing more warm, large exoplanets with nearby planets could help shed light on the nature of this second channel.

The \TESS pipeline \citep{jenk16,twic18,li18} recently discovered a pair of warm, large planet candidates orbiting \thisstar. Like the other systems in Fig.~\ref{fig:arch}, the putative planets are in or near an orbital resonance. Their proximity to resonance leads to detectable transit timing variations (TTVs). Based on expected \TESS planet yields, \citet{hadd18} predicted that significant mass constraints from TTVs would be possible for of order five planets. Here we seek to validate and characterize the \thisstar planet candidates and assess what additional follow-up is necessary to test theories for their origin. We characterize the host star in Section~\ref{sec:star}. In Section~\ref{sec:lc}, we describe our analysis of the \TESS data and extraction of planet parameters. We rule out most astrophysical false positive scenarios in Section~\ref{sec:valid}. We constrain the system's orbital architecture in Section~\ref{sec:arch} -- including mutual inclination, TTVs, eccentricities, and additional transit signals -- and the planets' masses sufficiently to confirm the planets. We present our conclusions in Section~\ref{sec:discuss}.

\section{Stellar Characteristics}
\label{sec:star}

\thisstar\ is an 11.5 \TESS apparent magnitude, main sequence K-dwarf. To better refine its parameters -- particularly the metallicity -- we obtained seven spectra of \thisstar with the ANU 2.3m Echelle spectrograph over a period of 11~days in 2018 Nov. These observations were also made to broadly constrain the mass of the planets and to check for obvious astrophysical false positive scenarios, such as line blending due to background stars. The ANU 2.3m/Echelle is located at Siding Spring Observatory, Australia. The spectrograph has a spectral resolution of $\lambda / \Delta \lambda \equiv R = 23000$, covering the wavelength region of 3900--6700~\AA. Observations are bracketed by ThAr arc lamp exposures for wavelength calibration. Instrument stability issues limit the radial velocities to a typical precision of only $\sim 500\,\mathrm{m\,s}^{-1}$ for this facility. Stellar parameters for \thisstar were derived using SpecMatch \citep{Yee:2017} on the ANU 2.3m/Echelle spectra, yielding atmospheric parameters of $T_\mathrm{eff} = 5045\pm110$~K, $\log g = 4.53\pm0.12$\,dex, and $\mathrm{[Fe/H]}=-0.16\pm0.09$\,dex. 

We use the approach described by \citet{daws15} to fit the observed stellar properties using the \citet{take07} and Dartmouth \citep{dott08} stellar evolution models. We perform an additional fit using the Dartmouth models to both the spectrum properties and the \emph{Gaia} DR2 parallax and apparent $g$ magnitude \citep{gaia16,gaia18}. We find that the measured atmospheric parameters are consistent with a main-sequence K-dwarf and list the derived stellar mass, radius, and density in column~2 of Table~\ref{tab:star}. The resulting values are in agreement with the \TESS Input Catalog (TIC; \citealt{stas18}) but more precise. We choose to use the Dartmouth values hereafter because the posteriors extend to a lower mass ($M_\star < 0.7 M_\odot$) than covered by the \citet{take07} models and because they allow us to fit the \emph{Gaia} DR2 parameters.

\begin{deluxetable*}{lcrrr}
 \tablecaption{Stellar Parameters\tablenotemark{a} for TOI-216 \label{tab:star}}
 \startdata
\\
Catalog Information \\
Parameters & Value & Source\\
 \hline
 ~~~~R.A. (h:m:s)                      &   04:55:55.3 & \emph{Gaia} DR2\\
~~~~Dec. (d:m:s)                      &  $-63$:16:36.2   & \emph{Gaia} DR2\\
~~~~Epoch							  &  2015.5         & \emph{Gaia} DR2 \\
~~~~Parallax (mas)                    & $5.59\pm0.03$ & \emph{Gaia} DR2\\
~~~~$\mu_{\mathrm{ra}}$ (mas yr$^{-1}$)        & $-22.7\pm0.04$  & \emph{Gaia} DR2 \\
~~~~$\mu_{\mathrm{dec}}$ (mas yr$^{-1}$)       & $-56.355\pm0.05$ & \emph{Gaia} DR2\\
~~~~$g$ magnitude & 12.163126\\
~~~~\emph{Gaia} DR2 ID                       &  4664811297844004352  &  \\
~~~~TIC ID                            & 55652896   & \\
~~~~TOI ID                            &  216    & \\
~~~~TIC \TESS magnitude & 11.504\\
~~~~$V$ magnitude\tablenotemark{b} & 12.393\\
\hline
\hline
Spectroscopic properties \\
Parameters & Spectrum & Takeda\tablenotemark{c} & Dartmouth\tablenotemark{d} & +\emph{Gaia}\tablenotemark{d}\\
\hline
~~~~Stellar effective temperature, $T_{\rm eff}$ [K] 							& 5045$\pm$110 		&50560$^{+1100}_{-1120}$ &50540$^{+1030}_{-1200}$& 50890$^{+430}_{-450}$\\
~~~~Iron abundance, [Fe/H]											&-0.16 $\pm$0.09		&-0.16$\pm 0.08$&-0.16$\pm0.09$&-0.15$^{+0.08}_{-0.09}$\\
~~~~Surface gravity, $\log g [$cm~s$^{-2}$] 							&4.53$\pm$0.12		&4.578$^{+0.02}_{-0.023}$ &4.58$^{+0.03}_{-0.04}$&4.58600$^{+0.003}_{-0.0350}$\\
~~~~Stellar mass, $M_{\star}$ [$M_{\odot}$]							&					&0.78$^{+0.04}_{-0.02}$ &  0.76$^{+0.04}_{-0.03}$ &  0.77$^{+0.03}_{-0.03}$ \\
~~~~Stellar radius, $R_{\star}$ [$R_{\odot}$] 							&					&0.765$^{+0.023}_{-0.02}$ & 0.74$^{+\pm0.043}_{-0.03}$& 0.747$^{+0.015}_{-0.014}$\\
~~~~Stellar density, $\rho_{\star}$ [$\rho_{\odot}$]						&					& 1.812$^{+0.14}_{-0.146}$	&  1.995$^{+0.213}_{-0.230}$ &	1.84$^{+0.14}_{-0.15}$	
 \enddata
 \tablenotetext{a}{As a summary statistic we report the median and 68.3\% confidence interval  of the posterior distribution.}
 \tablenotetext{b}{Using the relationship derived by \citet{jord10}, we compute the $V$ magnitude from the \emph{Gaia} $g$ magnitude and the Johnson-Cousins $I_C$ magnitude. We estimate the $I_C$ magnitude to be the \TESS magnitude, because the two band passes have the same center \citep{rick15}.}
\tablenotetext{c}{\citet{take07}}
\tablenotetext{d}{\citet{dott08}}
\end{deluxetable*}

\section{Light Curve Analysis}
\label{sec:lc}

\thisstar\ is located near the southern ecliptic pole, and is scheduled to be observed for 12 sectors of the first year of the \TESS Primary Mission. This paper is based on data from Sectors 1, 2, 3, 4, 5, and 6 (2018 July 25 -- 2019 January 7), during
which \thisstar\ was observed with CCD~1 on Camera~4, and from ground-based observatories.
\\
\\
\\
\subsection{Data from \TESS Mission}
We use the publicly available 2-min cadence data from the \TESS Alerts, which is processed with the Science Processing Operations Center pipeline. The pipeline, a descendant of the \Kepler mission pipeline based at the NASA Ames Research Center \citep{jenk02,jenk10,jenk16}, analyzes target pixel postage stamps that are obtained for pre-selected target stars. For \thisstar, the short cadence pipeline detected two threshold crossing event at periods 34.54~days and 17.1~days with high signal-to-noise. The candidates were also detected by the long cadence MIT Quick Look Pipeline \citep{sha19}.

\subsection{Ground-based Photometric Follow-up}

{ We used the resources of the \TESS Follow-up Observing Program (TFOP) Working Group (WG) Sub Group~1 (SG1)\footnote{\url{https://tess.mit.edu/followup/}} to collect seeing-limited time-series photometric follow-up of TOI-216.} The transit depths of both TOI-216 planet candidates, as predicted by the \TESS light curves, are deep enough to detect from the ground at high significance. Therefore our primary goal was to attempt to detect the transits using our higher spatial resolution ground-based imaging and a photometric aperture that is small enough to exclude the flux from known nearby stars that are bright enough to cause the \TESS detected events. The secondary goal was to identify or rule out potential nearby eclipsing binaries (Section~\ref{sec:valid}). 

We used the {\tt TESS Transit Finder}, which is a customized version of the {\tt Tapir} software package \citep{Jensen:2013}, to schedule photometric time-series follow-up observations. We initially scheduled observations for both planet candidates according to the public linear ephemerides derived from Sectors 1 and 2 \TESS data. Our eight time-series follow-up observations are listed in Table~\ref{tab:ground}. We used the AstroImageJ software package \citep{Collins2017} for data reduction and aperture photometry for all of our follow-up photometric observations. The facilities used to collect the TOI-216 observations are: Las Cumbres Observatory (LCO) telescope network \citep{brown2013}; Hazelwood Observatory; the Myers-T50 Telescope; and El Sauce Observatory. All LCO 1~m telescopes are equipped with the Sinistro camera, with a 4k x 4k pixel Fairchild back illuminated CCD and a 26.5 x 26.5 arcmin FOV. The LCO 0.4~m telescopes are mounted with an SBIG STX6303 2048 x 3072 pixels CCD with a 19 x 29 arcmin FOV. Hazelwood is a private observatory with an f/8 Planewave Instruments CDK12 0.32 m telescope and an SBIG STT3200 2.2K$\times$1.5K CCD, giving a $20\arcmin\times13\arcmin$ field of view. The Myers-T50 is an f/6.8 PlaneWave Instruments CDK17 0.43 m Corrected Dall-Kirkham Astrograph telescope located at Siding Spring, Australia.  The camera is a Finger Lakes Instruments (FLI) ProLine Series PL4710 - E2V, giving a $15\farcm5\times15\farcm5$ field of view. El Sauce is a private observatory with a Planewave CDK14 0.36 m telescope on a MI500/750F fork mount. The camera is an SBIG STT1603-3 1.5K$\times$1.0K CCD, giving a $18\farcm5\times12\farcm3$ field of view.

We observed five transits of \thisstarout at three epochs and confirmed that the transit events occur on target using follow-up apertures with radius $\sim 6\arcsec$. We conducted five \thisstarinn observations at four transit epochs and ruled out the $\sim 4$ parts per thousand transit events at the public linear ephemeris. However, with the later addition of data from \TESS sectors 3 and 4 to the TTV analysis, we determined that the large TTV signal caused the transit events to egress before our follow-up observations started. We then observed an out-of-transit sequence that occurred just prior to the newly determined transit ingress time to help constrain the TTV model (since the time of transit was not observable from our available facilities).  
\clearpage
\begin{table*}
\footnotesize
\caption{Observation Log}             
\label{tab:ground}      
\centering 
\begin{tabular}{c l l c c c c c c c}  
\hline\hline       
\noalign{\smallskip}                  
\multirow{2}{*}{TOI-216} & Date & \multirow{2}{*}{Telescope}\tablenotemark{$\dag$} & \multirow{2}{*}{Filter} & ExpT & Exp & Dur. & Transit & Ap. Radius  & FWHM\\
& (UTC) &  &  & (sec) & (N) & (min) & expected coverage\tablenotemark{$\ddag$} & (arcsec) & (arcsec)   \\
\noalign{\smallskip} 
\hline  
\noalign{\smallskip}                  
\multirow{6}{*}{\inn}
& 2018-11-22${^\ddag}$ & LCO-SSO-0.4  & i$^\prime$ & 90  & 54  & 100 & Ingress+30$\%$ & 8.5 & 7.5 \\
& 2018-12-09${^\ddag}$ & Myers-T50    & Lum        & 60  & 200 & 240 & full           & 8.3 & 4.6 \\
& 2018-12-26${^\ddag}$ & LCO-SSO-1.0  & i$^\prime$ & 30  & 85  & 99  & Ingress+25$\%$ & 7.0 & 2.8 \\
& 2019-01-29${^\ddag}$ & LCO-SAAO-1.0 & r$^\prime$ & 100 & 97  & 225 & Full           & 9.3 & 2.4 \\
& 2019-01-29${^\ddag}$ & LCO-SAAO-1.0 & i$^\prime$ & 25  & 181 & 198 & Full           & 5.8 & 2.2 \\
& 2019-02-15           & LCO-SSO-1.0  & Zs         & 60  & 160 & 236 & Out-of-Transit & 4.7 & 2.0 \\
\hline
\noalign{\smallskip} 
\multirow{7}{*}{\out}
& 2018-12-16 & LCO-SAAO-1.0  & i$^\prime$ & 90  & 75  & 180 & Egress+60$\%$  & 5.8 & 2.5 \\
& 2018-12-16 & LCO-SAAO-1.0  & i$^\prime$ & 39  & 331 & 450 & Full           & 5.8 & 2.1 \\
& 2019-01-20 & Hazelwood-0.3 & g$^\prime$ & 240 & 101 & 449 & Egress+70$\%$  & 5.5 & 3.2 \\
& 2019-02-23 & LCO-SAAO-1.0  & Zs         & 60  & 148 & 212 & Out-of-Transit & 6.2 & 2.5 \\
& 2019-02-24 & LCO-CTIO-1.0  & Zs         & 60  & 150 & 213 & In-Transit     & 6.2 & 2.5 \\
& 2019-02-24 & El Sauce-0.36 & Rc         & 30  & 514 & 303 & Egress+90$\%$  & 5.9 & 3.7 \\
& 2019-02-24 & LCO-SSO-1.0   & Zs         & 60  & 81  & 117 & Out-of-Transit & 6.2 & 2.5 \\
\noalign{\smallskip}
\hline
\noalign{\smallskip}

\end{tabular}
\tablenotetext{$\dag$}{Telescopes: \\
             LCO-CTIO-1.0: Las Cumbres Observatory - Cerro Tololo Astronomical Observatory (1.0 m) \\
             LCO-SSO-1.0: Las Cumbres Observatory - Siding Spring (1.0 m) \\
             LCO-SAAO-1.0: Las Cumbres Observatory - South African Astronomical Observatory (1.0 m) \\
             LCO-SSO-0.4: Las Cumbres Observatory - Siding Spring (0.4 m) \\
             Myers-T50: Siding Spring Observatory - T50 (0.43 m)\\
             Hazelwood-0.3: Stockdale Private Observatory - Victoria, Australia (0.32 m) \\
             El Sauce-0.36: El Sauce Private Observatory - El Sauce, Chile (0.36 m) \\
             }
             
\tablenotetext{$\ddag$}{Observations did not detect a transit event because they were scheduled using the initial public \TESS linear ephemeris. The TTV offset from the linear ephemeris is now known to be larger than the time coverage of the observations.}
\end{table*}

\subsection{Light curve fits}

\begin{figure*}
\begin{center}
\includegraphics{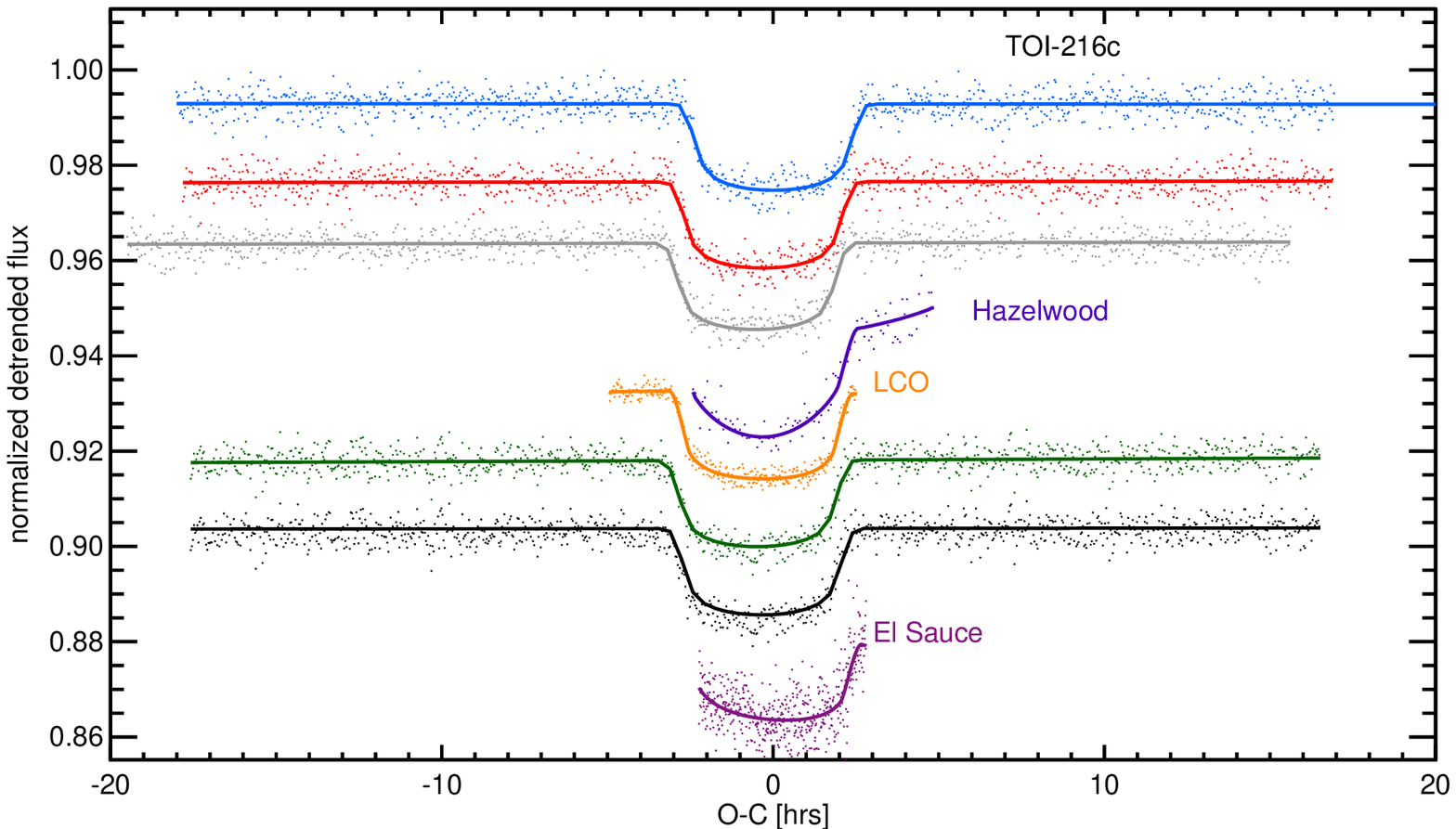}
\includegraphics{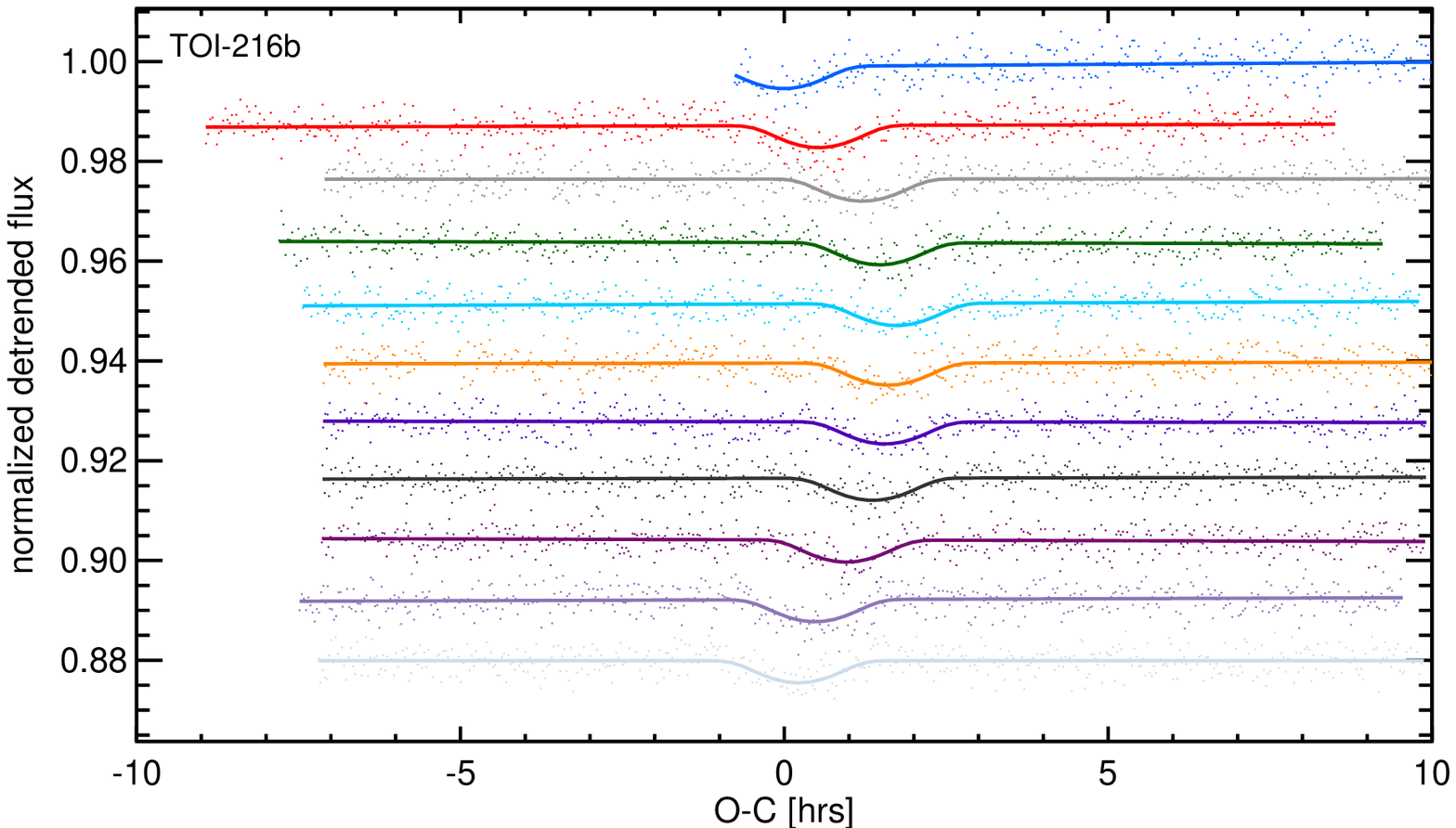}
\caption{
 \label{fig:transits} 
 Detrended light curves, color coded by transit epoch, spaced with arbitrary vertical  offsets, and with a model light curve overplotted. The light curves are phased based on a constant orbital period linear ephemeris to show the TTVs. }
\end{center}
\end{figure*}

We fit the transit light curves (Fig.~\ref{fig:transits}) using the TAP software \citep{gaza12}, which implements Markov Chain Monte Carlo using the \citet{mand02} transit model and the \citet{cart09} wavelet likelihood function, with the modifications described in \citet{daws14}. The results are summarized in Table~\ref{tab:216}. We use the presearch data conditioned (PDC) flux, which is corrected for systematic (e.g., instrumental) trends using cotrending basis vectors \citep{smit12,stum14}; the \citet{cart09} wavelet likelihood function (which assumes frequency$^{-1}$ noise) with free parameters for the amplitude of the red and white noise; and a linear trend fit simultaneously to each transit light curve segment with other transit parameters. We assign each instrument (\TESS, Hazelwood, LCO, El Sauce) its own set of limb darkening parameters because of the different wavebands. We use different noise parameters for \TESS, Hazelwood, LCO, and El Sauce. We adopt uniform priors on the planet-to-star radius ratio ($R_{p}/R_{\star}$), the log of the light curve stellar density $\rhocirc$ (i.e., equivalent the light curve parameter $d/R_\star$, where $d$ is the planet-star separation, converted to stellar density using the planet's orbital period and assuming a circular orbit), the impact parameter $b$ (which can be either negative or positive; we report $|b|$), the mid transit time, the limb darkening coefficients $q_1$ and $q_2$ \citep{kipp13}, and the slope and intercept of each transit segment's linear trend. For the Hazelwood, LCO, and El Sauce observations, we fit a linear trend to airmass instead of time.

\begin{deluxetable*}{rrl}
\tabletypesize{\footnotesize}
\tablecaption{Planet Parameters for \thisstarinn and \thisstarout Derived from the Light Curves \label{tab:216}}
\tablewidth{0pt}
\tablehead{
\colhead{Parameter}    & \colhead{Value\tablenotemark{a}}}
\startdata
\hline
\thisstarinn\\
Planet-to-star radius ratio, $R_{p}/R_{\star}$   				&0.11 &$^{+0.04}_{-0.02}$ \\
Planet radius, $R_p$ [$R_\oplus$] & 8.6 & $^{+2.9}_{-1.9}$\\
Light curve stellar density, $\rhocirc$  [$\rho_\odot$]  					&1.13 &$^{+0.29}_{-0.19}$ \\
$a/R_\star$\tablenotemark{b} &29.1&$^{+2.3}_{-1.8}$\\
Impact parameter, $|b|$ 	&	0.99 &$^{+0.05}_{-0.04}$\\
Sky-plane inclination, $i_{\rm sky}$  $[^\circ]$& 88.0&$^{+0.2}_{-0.2}$\\
Mid-transit times & 1325.328 &$^{+0.003}_{-0.004}$\\
&1342.431 &$^{+0.003}_{-0.003}$\\ 
&1359.539 &$^{+0.003}_{-0.003}$\\
&1376.631 &$^{+0.003}_{-0.003}$\\
&1393.723 &$^{+0.003}_{-0.003}$\\
&1427.879 &$^{+0.003}_{-0.003}$\\
&1444.958 &$^{+0.003}_{-0.003}$\\
&1462.031 &$^{+0.003}_{-0.003}$\\
&1479.094 &$^{+0.003}_{-0.003}$\\
&1496.155 &$^{+0.003}_{-0.003}$\\
&1513.225 &$^{+0.003}_{-0.003}$\\
\hline
\thisstarout \\
Planet-to-star radius ratio, $R_{p}/R_{\star}$   				&0.1236 &$^{+0.0008}_{-0.0008}$ \\
Planet radius, $R_p$ [$R_\oplus$] & 10.2 & $^{+0.2}_{-0.2}$\\
Light curves stellar density, $\rhocirc$  [$\rho_\odot$]  					&1.75 &$^{+0.04}_{-0.06}$ \\
$a/R_\star$\tablenotemark{b} &53.8&$^{+0.4}_{-0.6}$\\
Impact parameter, $|b|$ 					&0.11 &$^{+0.09}_{-0.00}$ \\
Sky-plane inclination, $i_{\rm sky}$  $[^\circ]$ & 89.89&$^{+0.08}_{-0.10}$\\
Mid-transit times & 1331.2851 &$^{+0.0007}_{-0.0007}$\\
&1365.8245 &$^{+0.0007}_{-0.0007}$\\ 
&1400.3686 &$^{+0.0007}_{-0.0007}$\\ 
&1434.9227&$^{+0.0007}_{-0.0007}$\\ 
&1469.4773&$^{+0.0007}_{-0.0007}$\\   
LCO&1469.4781&$^{+0.0004}_{-0.0004}$\\   
Hazelwood&1504.037 &$^{+0.002}_{-0.002}$\\  
El Sauce&1538.5939 &$^{+0.0015}_{-0.0015}$\\  
\hline
Minimum mutual inclination $[^\circ]$ & $1.8$ &$^{+0.2}_{-0.2}$\\
\enddata
\tablenotetext{a}{As a summary statistic we report the median and 68.3\% confidence interval  of the posterior distribution.}
\tablenotetext{b}{If the planet's orbit is not circular, this corresponds to the average planet-star-separation during transit divided by the stellar radius.}
\end{deluxetable*}

\begin{deluxetable*}{rrlrlrlrl}
\tabletypesize{\footnotesize}
\tablecaption{Light Curve Parameters for the \thisstar system\label{tab:216star}}
\tablewidth{0pt}
\tablehead{
\colhead{Parameter\tablenotemark{a}}    & \colhead{\TESS}&&\colhead{El Sauce}&&\colhead{LCO}&&\colhead{Hazelwood}}
\startdata
Limb darkening coefficient, $q_{1}$ 						& $0.33$& $^{+0.12}_{-0.09}$ 	&$0.5$& $^{+0.2}_{-0.2}$&$0.52$& $^{+0.15}_{-0.12}$ 	&$0.50$& $^{+0.23}_{-0.15}$ 			\\
Limb darkening coefficient, $q_{2}$ 						& $0.32$&$^{+0.14}_{-0.11}$&$0.30$&$^{+0.26}_{-0.16}$&$0.21$&$^{+0.08}_{-0.08}$&	 	$0.7$&$^{+0.2}_{-0.2}$	\\
Red noise,  $\sigma_r$	[ppm]		& 3000 & $^{+800}_{-900}$  	&10000 & $^{+4000}_{-4000}$&1500 & $^{+1600}_{-1000}$		& 4000 & $^{+3000}_{-3000}$  		\\
White noise, $\sigma_w$	[ppm]		& 2367&$^{+17}_{-17}$ & 3140&$^{+80}_{-80}$ & 1060&$^{+40}_{-40}$ & 2450&$^{+190}_{-190}$\\
\enddata
\tablenotetext{a}{As a summary statistic we report the mode and 68.3\% confidence interval  of the posterior distribution.}
\end{deluxetable*}

\begin{figure}
\begin{center}
\includegraphics{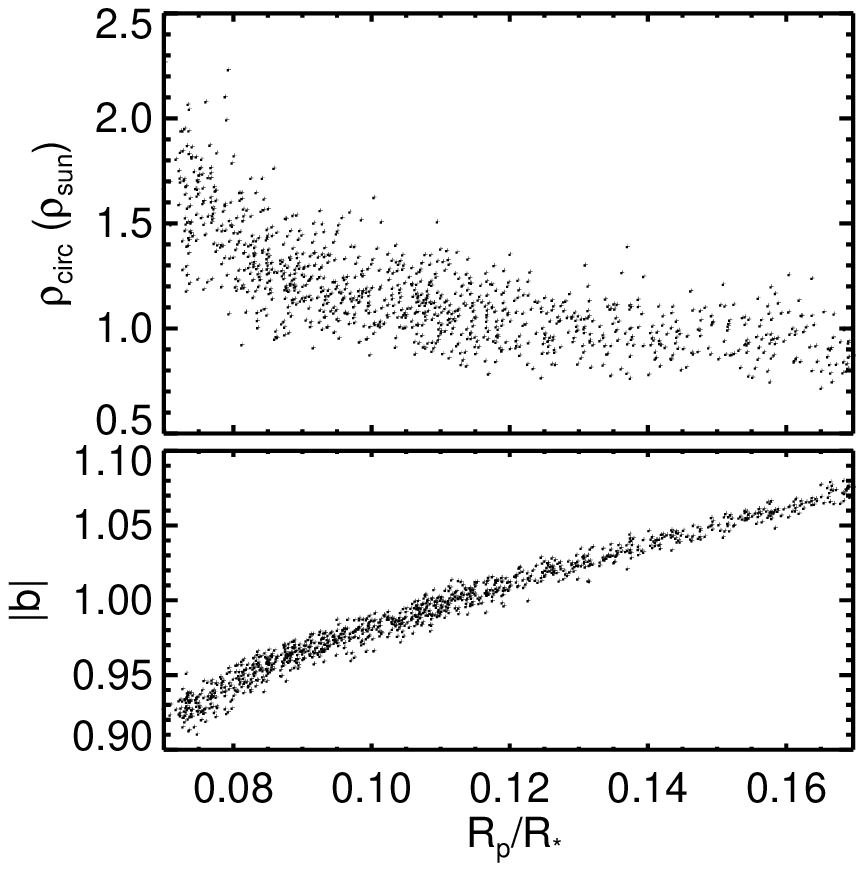}
\caption{
 \label{fig:rdr} 
Draws from the posterior distribution of correlated parameters $\rhocirc$, $R_p/R_\star$, and $|b|$ for TOI-216\inn, which has grazing transits. Larger $R_p/R_\star$ correspond to larger $|b|$ and smaller $\rhocirc$.
}
\end{center}
\end{figure}

The inner planet candidate's transits are grazing, so the planet-to-star radius ratio $R_p/R_\star$ is not well-constrained. We impose a uniform prior from 0 to 0.17, with the upper limit corresponding to a radius of 0.13 solar radii.  Fig.~\ref{fig:rdr} shows the covariance between $R_p/R_\star$ and the light curve stellar density $\rhocirc$ and impact parameter $b$. The larger the planet, the larger the impact parameter required to match the transit depth. The larger the impact parameter, the shorter the transit chord and the lower the light curve stellar density (which correlates with the transit speed) required to match the transit duration. Through its affect on $|b|$ and $\rhocirc$, the upper limit on $R_p/R_\star$ affects our inference of the inner planet's eccentricity and the mutual inclination between the planets; in Section~\ref{sec:arch}, we will assess the sensitivity to this upper limit.

\subsection{Search for additional transit signals}
\label{subsec:search}

We ran the box car least squares algorithm on the residuals of the light curve after removing the transit signal of \thisstarinn and \thisstarout. We used a duration of 2.5 hours, which corresponds to an impact parameter equal to planet \out's at an orbital period of 3~days. We did not find any signal with signal-to-noise larger than 7.3. Using per-point rms precision of 0.00233, this limit rules out any planets interior to \thisstarinn with a radius larger than 2.18 $R_{\oplus}$ or planets with periods less than 3~days and radii larger than 1.17 $R_{\oplus}$.  With future \TESS data from 12 sectors in total, the detection threshold for all planets interior to \thisstarinn will be lowered to 1.13 $R_{\oplus}$ planets.

\section{Validation}
\label{sec:valid}

Here we seek to validate the planet candidates by ruling out false positive scenarios using follow up observations and dynamical arguments. In Section~\ref{subsec:RV}, we consider and rule out unblended astrophysical false positive scenarios using radial velocity (RV) measurements. In \ref{subsec:phot}, we consider and rule out { most} blended false positive scenarios using photometry. We summarize the results in Section~\ref{subsec:summ}.

\subsection{Low precision radial-velocity follow up to rule out stellar companions to \thisstar}
\label{subsec:RV}

One or both transiting objects could be brown dwarf or stellar companions to \thisstar.  The following astrophysical false positive scenarios can be tested through radial velocity follow up: \thisstarinn and/or \thisstarout is a brown dwarf; \thisstarinn (which has a poorly constrained transit depth) is an unblended stellar companion; or \thisstarinn and/or \thisstarout is a blended stellar companion to \thisstar with a background or bound star diluting the transit depth.

{ If both objects are transiting \thisstar, but one or both is of brown dwarf or stellar mass, the system would be unstable if the objects are not in resonance or, if in resonance, the mass of the secondary would cause large TTVs incompatible with those observed (Section~\ref{sec:ttvs}). Furthermore, the brown dwarf scenario is less likely a priori.  \citet{grie17} find that the occurrence rate of brown dwarfs with orbital periods less than 300~days is about 0.56\%, compared to 4.0\% for planets $>0.3 M_{\rm Jup}$ \citep{cumm08}.}

We use radial velocity (RV) measurements to put mass limits on any companion to \thisstar. The spectra described in Section~\ref{sec:star} shows no large radial velocity variations, with the measurements exhibiting a scatter of $470\,\mathrm{m\,s}^{-1}$. From these velocities, we derive the $3\sigma$ upper limit on the masses of the inner planet to be $\sim 18\,M_J$ and the outer planet to be $\sim 25\,M_J$. The upper limits rule out any scenario involving a stellar companion to \thisstar. The constraints also support our limit on $R_p/R_\star$ for the light curve fits for \thisstarinn (Section~\ref{sec:lc}) corresponding to 1.3 $R_J$ because radii only start to increase above $\sim 1 R_J$ at around $60 M_J$ (e.g., \citealt{hatz15}, Fig.~2). The scenario in which one or both objects are brown dwarf companions is not ruled out by the RVs but will be ruled out by the TTVs in Section~\ref{sec:ttvs}.

\subsection{Photometry rules out most blended false~positive~scenarios}
\label{subsec:phot}

Analysis of systems with multiple transiting planet candidates from \Kepler has shown that the transit-like events have a higher probability of being caused by bona fide planets \cite[e.g.,][]{Lissauer2012} compared to single-planet candidate systems, lending credibility to the planetary nature of the transit-like events associated with TOI-216. However, the pixel scale of \TESS is larger than \Kepler's (${21\arcsec}$ for \TESS vs. ${4\arcsec}$ for \Kepler) and the point spread function of \TESS could be as large as ${1\arcmin}$, both of which increase the probability of contamination of the \TESS aperture by an nearby eclipsing binary. For example, a deep eclipse in a nearby faint eclipsing binary might cause a shallow transit-like detection by \TESS on the target star due to the dilutive effect of blending in the \TESS aperture. 

A scenario in which both \thisstarinn and \thisstarout are orbiting the same background binary is ruled out by the TTVs (Section~\ref{sec:ttvs}). One object could be a planet-mass companion to \thisstar\ and the other a background binary. Alternatively, both objects could be background binaries.

From a single sector of \TESS data, the one standard deviation centroid measurement uncertainty is $2\farcs58$ for \thisstarinn and $3\farcs3$ for \thisstarout. \thisstarout would need to fully eclipse a star with Tmag 15.85 to cause the blend, and \thisstarinn would need to fully eclipse a star with Tmag 17.5 to cause the blend. The brightest \emph{Gaia} D2 object within $40\arcsec$ has \emph{Gaia} $rp$ magnitude of 16.8 and is $3\farcs768$ away and therefore is marginally compatible with a blend scenario for \thisstarinn. The second brightest \emph{Gaia} object within $40\arcsec$ has \emph{Gaia} $rp$ magnitude of 17.94, which cannot cause either of the transit signals we see. 

We use higher spatial resolution ground-based time-series imaging to attempt to detect the transit-like events on target and/or to identify or rule out potential nearby ecliping binaries out to ${2.5\arcmin}$ from TOI-216. The higher spatial resolution and smaller point spread function of the ground-based observations facilitates the use of much smaller photometric apertures compared to the \TESS aperture to isolate a possible transit or eclipse signal to within a few arcseconds of the center of the follow-up aperture. From the ground, follow-up apertures exclude the flux of all known neighboring stars, except the two $\sim4\arcsec$ \emph{Gaia} DR2 neighbors. We collected observations of \thisstarout in both g$^\prime$ and i$^\prime$ filters (Section~\ref{sec:lc}) and found no obvious filter-dependent transit depth, which strengthens the case for a planetary system. 

\subsection{Validation summary}
\label{subsec:summ}
{ In summary, we can rule out all astrophysical false positive scenarios with a couple exceptions. First, \thisstarinn could be a blended binary orbiting the 16.8 $rp$ magnitude \emph{Gaia} DR2 object, in which case it would need a 53\% transit depth. Second, \thisstarinn and/or \thisstarout could be a binary orbiting a star located at the same sky position as \thisstar, creating a blend not resolved by \emph{Gaia}.} However, we will show in Section~\ref{sec:ttvs} that the two transiting objects are fully compatible with causing each other's TTVs and that the TTVs have concavity in opposite direction (i.e., one planet loses orbital energy as the other gains). This false positive scenario would require the extremely unlikely configuration in which both objects happen to have an orbital period ratio near 2:1, happen to have non-transiting companions in or near orbital resonance causing their TTVs, and the TTVs happen to have opposite sign. Therefore we consider the system to be validated.
\\
\\
\section{Orbital Architecture}
\label{sec:arch}

Here we explore the orbital architecture of the \thisstar system through analysis of the transiting timing variations (TTVs), transit shape and duration, and limits on additional transiting planets.

\subsection{TTV overview}
\begin{figure}
\begin{center}
\includegraphics{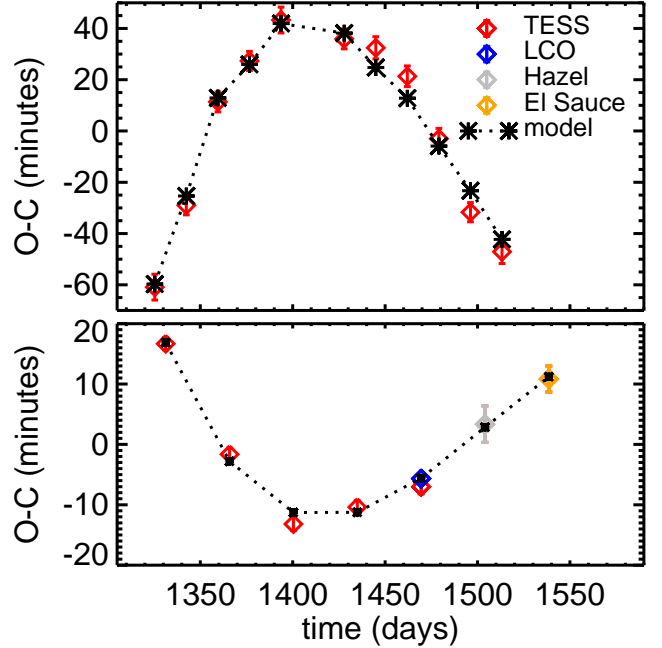}
\caption{
 \label{fig:oc} 
Observed mid-transit times (diamonds) with subtracted best-fit linear ephemeris for TOI-216\inn (top) and TOI-216\out (bottom), with the best fit model overplotted (asterisks, dotted line).
}
\end{center}
\end{figure}

{
Both candidates exhibit significant deviations from a linear transit time ephemeris (Fig.~\ref{fig:oc}), evidence for their mutual gravitational perturbations. These transit timing variations (TTVs) occur on two timescales. The first is the synodic timescale, $\tau_{\rm syn} = P_\out/(P_\out/P_\inn-1)$, which is the interval of time between successive planetary conjunctions. 
The second -- for planets near the 2:1 resonance -- is the super-period\footnote{The super-period may be longer or shorter for planets in orbital resonance experiencing fast precession.}, $\tau_{\rm s-p}\approx|P_\out/(2-P_\out/P_\inn)|$, the timescale over which the planets have their conjunctions at the same longitude; $\tau_{\rm s-p}$ depends on the proximity of the ratio of the orbital periods to 2. 

The synodic TTV signal, known as the chopping effect because it produces a saw-tooth like pattern (see \citealt{deck15} and references therein), depends on the perturbing planet's mass, which determines the strength of the kick at conjunction. To first order, the chopping effect does not depend on eccentricity. 

The super-period TTV signal, known as the near-resonant effect (e.g., \citealt{lith12}), has a sinusoidal shape. The near-resonant effect generates a forced eccentricity for each planet, and the free eccentricity is an extra component that contributes to the total eccentricity. The near-resonant TTV amplitude depends on the perturbing planet's mass and the free eccentricity of the transiting and perturbing planets. To first order, the ratio of near-resonant signal amplitudes depends only on the planets' mass ratio (e.g., \citealt{lith12}'s Eqn. 14--15). Therefore TTVs covering a significant fraction of the super-period can provide a good estimate of the mass ratio.

For planets near resonance, the amplitude of the near-resonant effect is typically much larger than the amplitude of the chopping effect. Measuring the chopping and near-resonant signals for both transiting planets -- assuming there are no additional planets in the system contributing significantly to the TTVs -- would allow us to uniquely constrain their masses and eccentricities. 
}

\subsection{Evidence for free eccentricity}

\begin{figure*}
\begin{center}
 \includegraphics{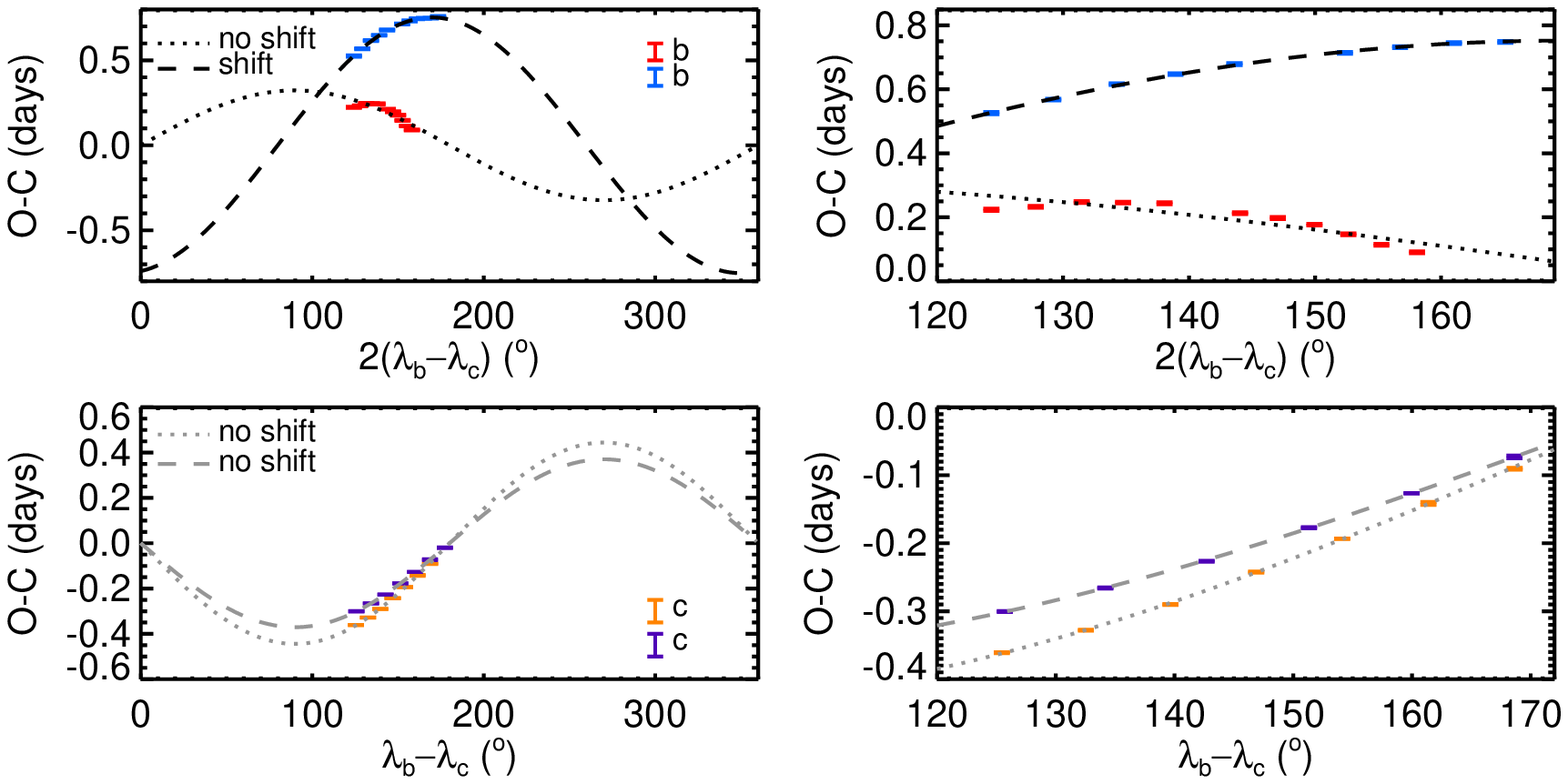}
 \includegraphics{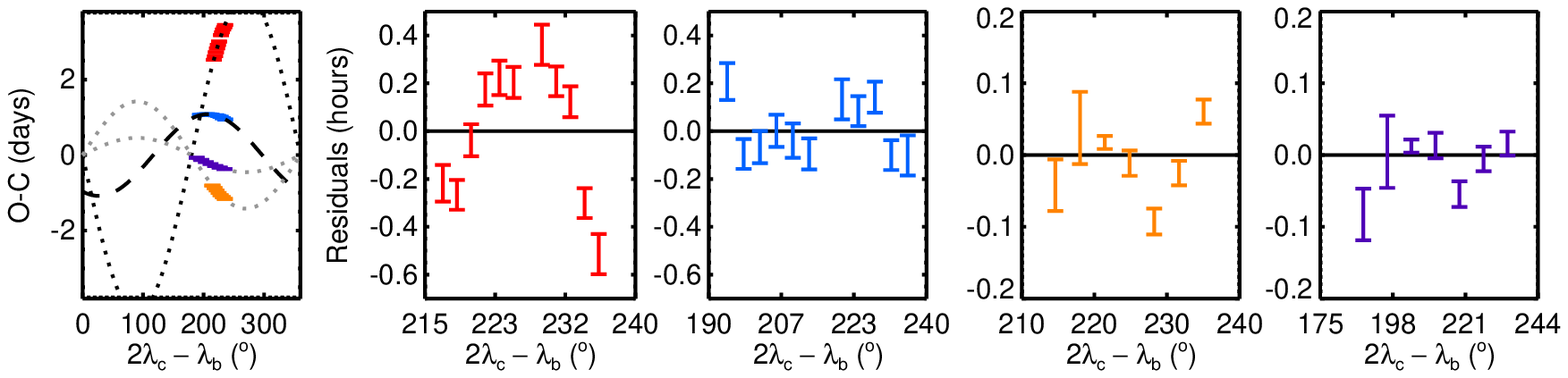}
\caption{
 \label{fig:oc_phase} Evidence for free eccentricity from TTVs plotted as function of phase, where $\lambda_i$ is the mean longitude of the $i^{\rm th}$ planet's orbit. Top row: Inner planet's TTVs. We plot the best fit non-phase shifted sinusoid as a dotted line that goes with the red observed points, and the best fit phase-shifted sinusoid as a dashed line that goes with the blue points. Note that the red and blue points are different because orbital period and first transit epoch are also free parameters. A non-phase shifted sinusoid is inconsistent with the observed TTVs of \thisstarinn, so we infer the planets have free eccentricity. row~2: Outer planet's TTVs. No phase shift in $\lambda_\inn-\lambda_\out$ is necessary. The linestyle corresponds to the same linear ephemeris as used in row~1. The orange (purple) points use the same linear ephemeris as the red (blue) points in row~1. Bottom row: TTVs phased to $2\lambda_\out-\lambda_\inn$. For the inner planet, a phase shift is necessary to match the inner planet's observed TTVs (i.e., the red points are not well-fit by the model).
}
\end{center}
\end{figure*}

The phasing of the TTVs allows to diagnose at least one planet likely has significant free eccentricity. In Fig.~\ref{fig:oc_phase} we plot the TTVs as a function of  phase. The top panel shows the TTVs of the inner planet phased with $2(\lambda_\inn-\lambda_\out$), where $\lambda$ is the mean longitude (Section \ref{subsec:fits}). If the free eccentricities are zero, the TTVs should follow a sinusoid with no phase shift \citep{deck15}. The non-phase shifted sinusoid is inconsistent with the observed TTVs of \thisstarinn, so we infer at least one planet has free eccentricity. [For the outer planet, no phase shift in $\lambda_\inn-\lambda_\out$ (Fig.~\ref{fig:oc_phase}, row~2) is necessary.] We also follow \citet{lith12} and plot the TTVs phased to $2\lambda_\out-\lambda_\inn$ (Fig.~\ref{fig:oc_phase}, row~3; equivalent to rows~1--2 because transit times are sampled at the planets' orbital period) and find that again a phase shift is necessary to match the inner planet's observed TTVs, indicating free eccentricity for one or both planets. 

The chopping signal would appear as additional harmonics, i.e., $\lambda_\inn-\lambda_\out$, $3(\lambda_\inn-\lambda_\out)$, etc. for \thisstarinn and $2(\lambda_\inn-\lambda_\out)$ and $3(\lambda_\inn-\lambda_\out)$, etc. for \thisstarout. The fact that a sinusoid goes through the data points in Fig.~\ref{fig:oc} without these additional harmonics gives us a sense that the chopping signal will not be easily measured in this dataset. There will be a degeneracy between planet masses and free eccentricity.

\subsection{A large range of best-fit planet masses}
\label{subsec:fits}
We fit the transit times using our N-body TTV integrator model \citep{daws14}. Our model contains five parameters for each planet: the mass $M$, orbital period $P$, mean longitude at epoch $\lambda$, eccentricity $e$, and argument of periapse $\omega$. For each planet, we fix the sky plane inclination $i_{\rm sky}$ to the value in Table~\ref{tab:216} and set the longitude of ascending node on the sky to $\Omega_{\rm sky}=0$. We use the conventional coordinate system where the $X-Y$ plane is the sky plane and the $Z$ axis points toward the observer. See \citet{murr10} for a helpful pedagogical description of the orbital elements. 

\begin{figure}
\begin{center}
\includegraphics{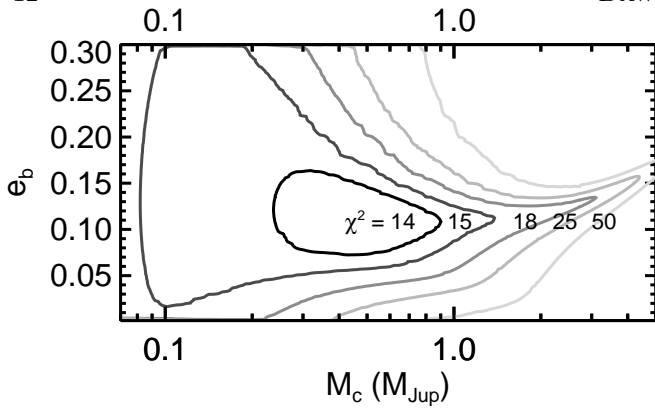}
\caption{
 \label{fig:contour} 
Contours of $\chi^2$ show degeneracy between the inner planet's (osculating) eccentricity and the outer planet's mass. The best-fit solutions occupy the innermost contour.
}
\end{center}
\end{figure}

To explore the degeneracy between mass and eccentricity, we use the Levenberg-Marquardt alogrithm implemented in IDL {\tt mpfit} \citep{mark09} to minimize the $\chi^2$ on a grid of $(M_\out, e_\inn)$. { We report the total $\chi^2$ for eighteen transit times and ten free parameters, i.e., eight degrees of freedom.} The resulting contour plot is shown in Fig.~\ref{fig:contour}. The lowest $\chi^2$ fits, i.e., those with $13 < \chi^2 < 18$, are possible for a range of outer planet masses ($M_\out<3.0\,M_{\rm Jup}$). However, for small outer planet masses, a large range of inner planet eccentricities allow for a good fit, whereas a particular value of the eccentricity ($e_\inn \sim 0.13$ is necessary for larger planet masses. (See also discussions by \citealt{hadd17} and \citealt{miga18}.) Because there is so much more ``real estate'' in parameter space at low outer planet masses, a Markov Chain Monte Carlo (MCMC) will identify this type of solution as most probable. However, if we have a priori reason to suspect the outer planet is massive -- like a large transit depth -- and/or that free eccentricities are low, we could be misled. 

\begin{figure}
\begin{center}
\includegraphics{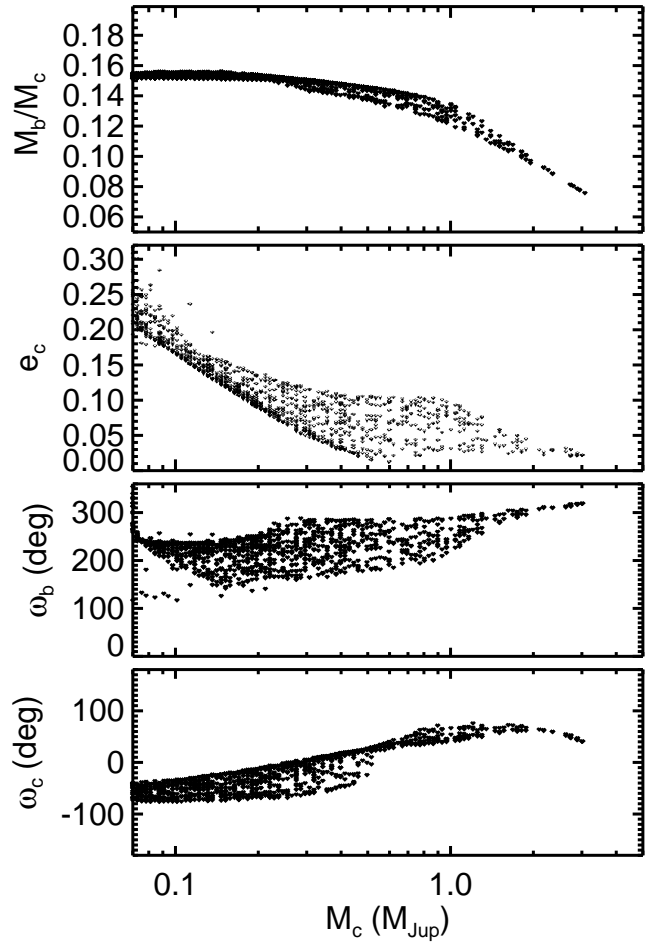}
\caption{
 \label{fig:others} 
Correlations between parameters in best fit solutions ( $\chi^2 < 18$).  Larger outer planet masses correspond to smaller mass ratios ($M_\inn/M_\out$) and smaller inner planet eccentricities; outer planet mass maps to particular ranges of the argument of periapse $\omega$.
}
\end{center}
\end{figure}

Fig.~\ref{fig:others} shows how other parameters correlate with the outer planet's mass $M_\out$. The mass ratio, $M_\inn/M_\out$, of the planets is about 0.17 for $M_\out < 0.5 M_{\rm Jup}$ and decreases for larger $M_\out$. Solutions with $M_\out < 0.5 M_{\rm Jup}$ have larger values for the eccentricity of planet \out. (Note that the eccentricity plotted in Fig.~\ref{fig:contour} and \ref{fig:others} is the osculating eccentricity; we will explore how these solutions translate to free and forced eccentricities in Section~\ref{subsec:long}.) The arguments of periapse $\omega_b$ and $\omega_c$ for planets \inn and \out also correlate with planet \out's mass.

\subsection{Longterm behavior of best-fit solutions}
\label{subsec:long}

\begin{figure}
\begin{center}
\includegraphics{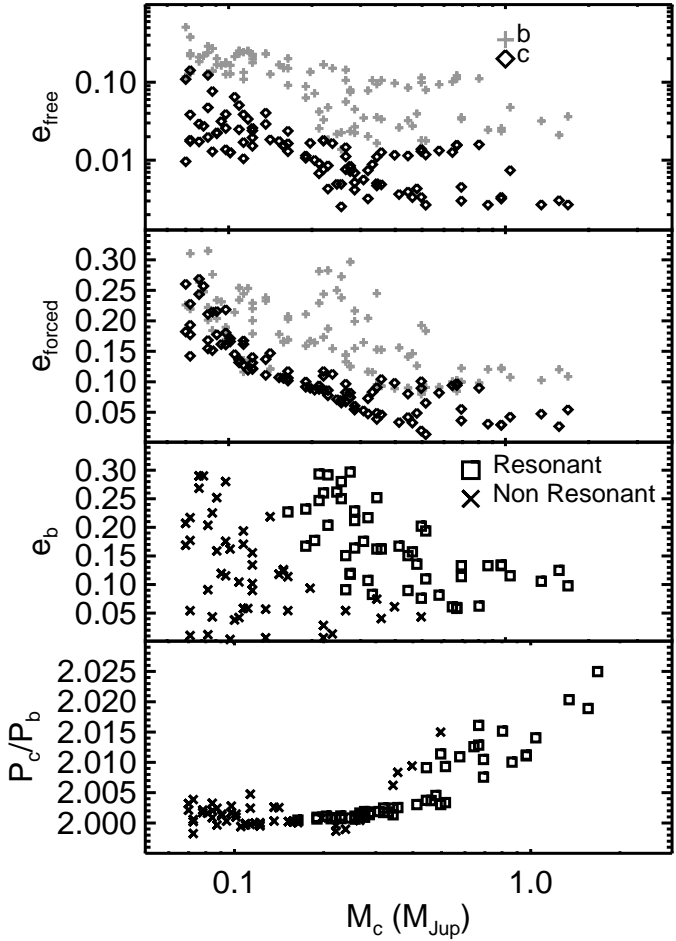}
\caption{
 \label{fig:long} 
Long-term ($10^6$~days) behavior of solutions with $\chi^2<11$: free eccentricity (row~1; calculated as the maximum deviation from the median eccentricity), forced eccentricity (row~2; calculated as the median eccentricity); $e_b$ and orbital resonance (row~3); and time-averaged orbital period ratio (row~4).
}
\end{center}
\end{figure}

We integrate the $\chi^2 < 18$ solutions for $10^6$~days using {\tt mercury6} \citep{cham96} to assess the longer term behavior (Fig.~\ref{fig:long}). We find that resonant argument $2\lambda_\out-\lambda_\inn-\varpi_\inn$ librates for the high $M_\out$ ($M_\out\gtrapprox0.3\,M_{\rm Jup}$) solutions but not for the lower $M_\out$ solutions. Larger $M_\out$ solutions have lower free and forced eccentricities for both planets (Fig.~\ref{fig:long}, rows~2--3). Period ratios $P_\out/P_\inn$ are wider of the 2:1 for the higher $M_\out$ solutions. We extend the simulations to 10 Myr and find that all configurations remain stable over that interval.

\subsection{Transit exclusion intervals}

We use ground-based observations in which an ingress or egress for \thisstarinn is excluded (Table~\ref{tab:ground}) to check solutions. Before the \TESS Sector 6 data were available, the exclusion interval on the December 26 observation ruled out some solutions. Almost all solutions based on Sector 1--6 are consistent with no ingress or egress during the intervals in Table~\ref{tab:ground}.
 
\subsection{Ruling out the lowest-mass solutions with the ``photoeccentric'' effect}

\begin{figure}
\begin{center}
 \includegraphics{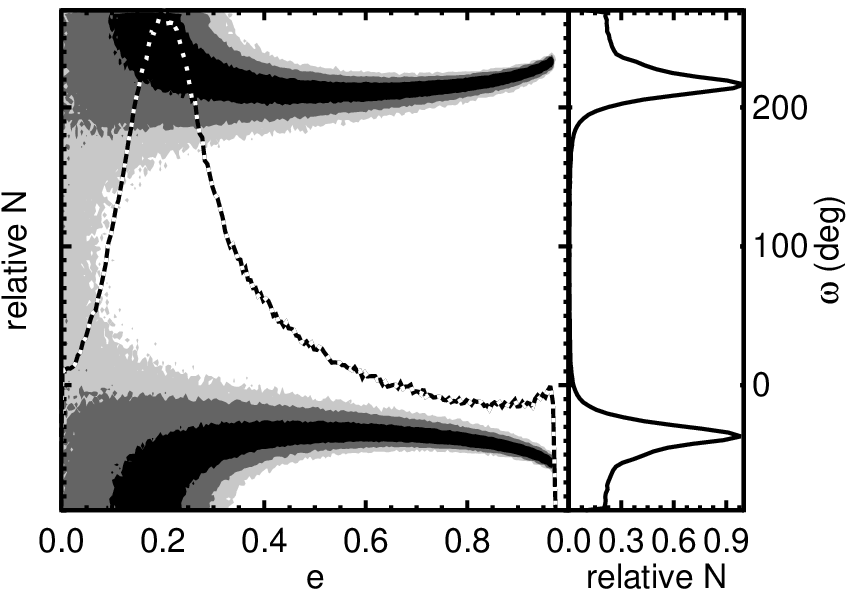}\\
  \includegraphics{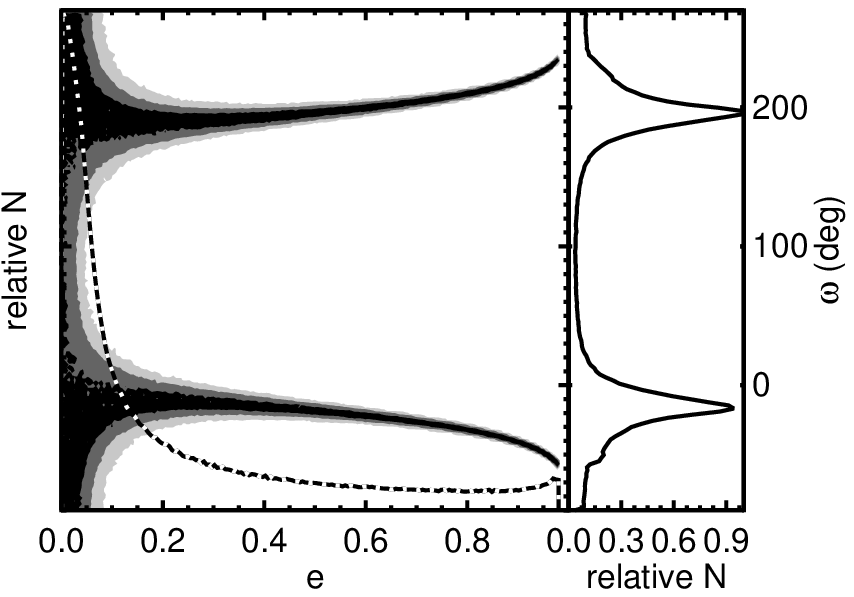}
\caption{
 \label{fig:ecc} 
 Joint posterior, $\omega$ vs. $e$, for TOI-216\inn (top) and TOI-216\out (bottom). The black (gray, light gray) contours represent the \{68.3,95,99\}\% probability density levels (i.e., 68\% of the  posterior  is  contained  within  the  black  contour).  Overplotted  as  a  black and white dotted line is a histogram of the eccentricity posterior probability distribution marginalized over $\omega$. The transit shapes and durations are consistent with low eccentricity orbits, but moderately eccentric orbits are not ruled out for special ellipse orientations that result in similar planet-star separations to the circular case.
}
\end{center}
\end{figure}

The light curve stellar densities (Table~\ref{tab:216}) are similar to the true stellar density (Table~\ref{tab:star}), consistent with the planets being on nearly circular orbits. We follow \citet{daws12} to estimate the candidates' eccentricities from the light curve using the ``photoeccentric effect,'' but instead of applying the approximations appropriate for a grazing transit, we use the full Eqn. 15 from \citet{kipp10}. We find eccentricities that could be low for both candidates; their modes and 68.3\% confidence intervals a are: $e_\inn=0.20_{-0.06}^{+0.48}, e_\out = 0.025_{-0.004}^{+0.490}$ (Fig.~\ref{fig:ecc}). The medians and their 68.3\% confidence intervals are $e_\inn=0.30_{-0.16}^{+0.38}, e_\out = 0.10_{-0.08}^{+0.41}$. High eccentricities are not ruled out, e.g., the posterior probability of $e>0.5$ is 28\% for \thisstarinn and 16\% \thisstar\out. The posterior probability of an eccentricity less than 0.01 is 0.7\% for \thisstarinn and 8\% for \thisstarout. 

\begin{figure}
\begin{center}
\includegraphics{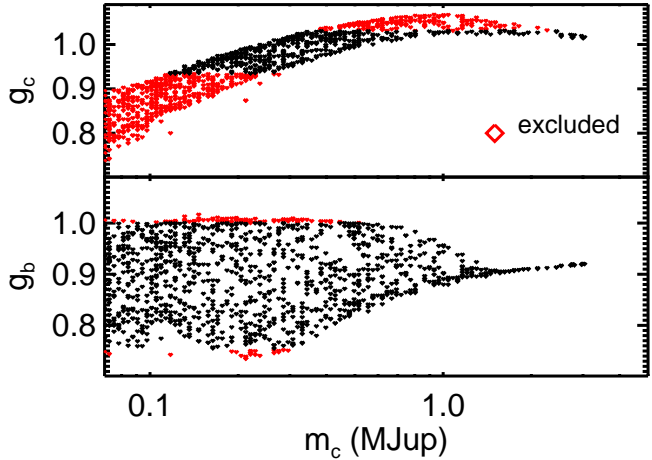}
\caption{
 \label{fig:exclude} 
Constraints on $g=\left(\rho_{\rm circ}/\rho_\star\right)^{1/3}$ from the light curve rule out a subset of solutions (red; inconsistent with $g$ outside the 2.5--97.5 percentile). Solutions with $\chi^2 < 18$ are plotted.
}
\end{center}
\end{figure}

The constraints on the eccentricity from the light curve allow us to rule out the lowest-mass solutions (Fig.~\ref{fig:exclude}). These solutions -- which correspond to an eccentric \thisstarout with its apoapse near our line of sight -- would produce a transit duration that is too long. Some higher-mass solutions that correspond to an eccentric  \thisstarout with its periapse near our line of sight are also ruled out. 

\subsection{MCMC fits}
Following \citet{daws14}, we derive posteriors for the parameters using Markov Chain Monte Carlo with the Metropolis-Hastings algorithm. We incorporate the transit exclusion intervals and light curve stellar density (i.e., combining the $\rhocirc$ posterior from the light curve and $\rho_\star$ posterior from the Dartmouth models) into the MCMC. Instead of including the orbital period and mean longitude at epoch as parameters in the MCMC, we optimize them at each jump using the Levenberg-Mardquardt algorithm. We visually inspect each parameter for convergence.

We perform two fits with different priors to explore  both ends of the parameter degeneracy
  evident in the grid of outer mass vs. inner eccentricity (Fig.~\ref{fig:contour}). The first solution (Table~\ref{tab:ttv216}, column~1) imposes { uniform} priors on eccentricities and { log} uniform priors on mass { (i.e., priors that are uniform in log space)}; the second (Table~\ref{tab:ttv216}, column~2) imposes uniform priors on { mass} and { sets $e_c=0$ (which we found to yield indistinguishable results from an eccentricity prior that is uniform in log space)}. All other fitted parameters (orbital period, mean longitude, argument of periapse) have uniform priors.  The uniform priors on mass favors the higher-mass solutions seen in Fig.~6--8, whereas the log uniform prior on mass favor the lower-mass solutions. 

Because the results are so prior-dependent (every parameter in Table~\ref{tab:ttv216} differs significantly between the two solutions except \thisstarinn's eccentricity of $\sim 0.2$), we do not recommend currently adopting either solution. Instead, the MCMC approach is a way to formally separate the two types of solutions seen in the grid search and to incorporate the light curve stellar densitities and transit exclusion windows into the likelihood function.

\begin{deluxetable}{rrlll}
 \tablecaption{Planet Parameters for \thisstarinn and \thisstarout Derived from TTVs \label{tab:ttv216} }
 \tablehead{
 \colhead{Parameter}    & \colhead{Soln 1\tablenotemark{a,b}}&
 & \colhead{Soln 2\tablenotemark{a,c}}}
 \startdata
$M_\inn$ ($M_{\rm Jup}$) & 0.05 &$^{+0.023}_{-0.03}$ & 0.10 &$^{+0.03}_{-0.02}$\\
$M_\inn/M_\out$ & 0.149 & $^{+0.011}_{-0.012}$& 0.133 & $^{+0.010}_{-0.010}$\\
$e_\inn$ & 0.214 & $^{+0.154}_{-0.048}$ & 0.15 & $^{+0.04}_{-0.03}$ \\
$\varpi_\inn$ (deg.) & 240&$^{+40}_{-30}$& 293&$^{+7}_{-10}$\\
$M_\out$ ($M_{\rm Jup}$) & 0.26 &$^{+0.14}_{-0.17}$& 0.57 &$^{+0.21}_{-0.16}$\\
$e_\out$ & 0.06 & $^{+0.11}_{-0.03}$& \\
$\varpi_\out$ (deg.) &-30&$^{+30}_{-60}$\\
$\Delta \varpi$  (deg.)& -80&$^{+30}_{-30}$  \\
$2\lambda_\out - \lambda_\inn -\varpi_\out$ (deg).&-20&$^{+40}_{-30}$\\
$2\lambda_\out - \lambda_\inn -\varpi_\inn$ (deg).&60&$^{+11}_{-14}$&41&$^{+7}_{-6}$\\
\enddata
 \tablenotetext{a}{As a summary statistic we report the median and 68.3\% confidence interval of the posterior distribution.}
  \tablenotetext{b}{Uniform prior on eccentricity and log uniform prior on mass}
   \tablenotetext{c}{{ $e_c=0$} and uniform prior on mass.}
\end{deluxetable}

\subsection{Mass-radius}
\label{sec:ttvs}

\begin{figure}
\begin{center}
\includegraphics{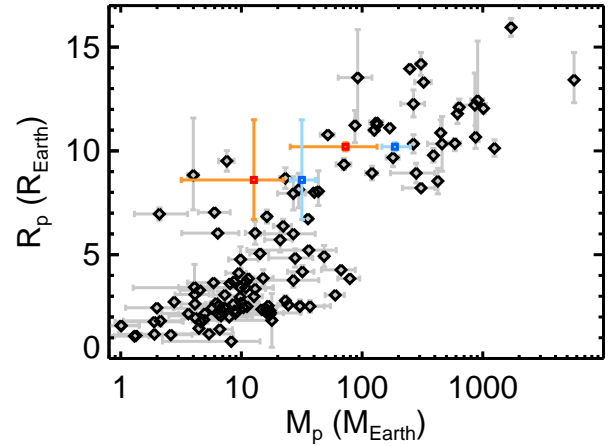}
\caption{\label{fig:rm} 
Warm (10--200~day orbital period) planets with both mass and radius measurements (exoplanets.eu), including TOI-216 (red, Solution 1; blue, Solution 2). 
}
\end{center}
\end{figure}

 We plot the two solutions on a mass-radius plot in Fig.~\ref{fig:rm}. \thisstarout's radius is comparable to other known exoplanets for both mass solutions. The same is true for \thisstarinn if its radius is close to the lower limit derived from its grazing transits. However, if its radius is somewhat larger than the lower limit, the lower-mass solution would correspond to a very low density.
 
\subsection{Predictions for future transits}
\label{subsec:predict}
In Table~\ref{tab:future}, we tabulate the predicted times for { missed and} future transits. { For the inner planet, the predictions of the two solutions overlap within one standard deviation for each transit. However, the outer planet's transits differ between the solutions, so the next few sectors of \TESS data may help distinguish between them.} 

\begin{deluxetable}{rrlll}
 \tablecaption{{ Missed and} future transit times\label{tab:future}}
 \tablehead{
 \colhead{Solution 1\tablenotemark{a,b}}&
 & \colhead{Solution 2\tablenotemark{a,c}}}
 \startdata
 \thisstarinn\\
1530.286&$^{+0.006}_{-0.004}$ & 1530.295&$^{+0.011}_{-0.007}$ \\
1547.351&$^{+0.009}_{-0.007}$& 1547.363&$^{+0.013}_{-0.010}$ \\
1564.413&$^{+0.013}_{-0.010}$ & 1564.430&$^{+0.020}_{-0.015}$  \\
1581.479&$^{+0.019}_{-0.014}$  & 1581.50&$^{+0.02}_{-0.02}$ \\
1598.54&$^{+0.03}_{-0.02}$ & 1598.58&$^{+0.03}_{-0.03}$ \\
1615.61&$^{+0.04}_{-0.02}$ & 1615.65&$^{+0.04}_{-0.04}$ \\
 \thisstarout\\
1573.09&$^{+0.03}_{-0.03}$& 1573.16&$^{+0.04}_{-0.03}$ \\
1607.63&$^{+0.04}_{-0.04}$& 1607.71&$^{+0.05}_{-0.04}$ \\
1642.18&$^{+0.04}_{-0.05}$ & 1642.26&$^{+0.05}_{-0.04}$\\
1676.72&$^{+0.05}_{-0.05}$ & 1676.82&$^{+0.06}_{-0.05}$\\
1711.26&$^{+0.05}_{-0.06}$ & 1711.37&$^{+0.07}_{-0.05}$\\
1745.81&$^{+0.06}_{-0.06}$ & 1745.92&$^{+0.07}_{-0.06}$\\
\enddata
 \tablenotetext{a}{As a summary statistic we report the { median} and 68.3\% confidence interval  of the posterior distribution.}
  \tablenotetext{b}{Uniform prior on eccentricity and log uniform prior on mass}
   \tablenotetext{c}{Log uniform prior on eccentricity and uniform prior on mass.}
\end{deluxetable}

\subsection{Mutual inclination}

A larger impact parameter for an inner planet than an outer planet points to at least a small mutual inclination between their orbits. The difference in the \thisstarinn and \thisstarout's sky plane inclination (Table~\ref{tab:216}) corresponds to a mutual inclination of at least $1^\circ.90^{+0.15}_{-0.34}$ (mode; the median is $1^\circ.8^{+0.2}_{-0.3}$). This value is a minimum because we do not know the component of the mutual inclination parallel to the sky plane. { Future observations of transit duration variations -- and depth changes for the grazing transit -- may allow for constraints on the full 3D orbital architecture.}

\subsection{Comparison to other work}
{ While this manuscript was in preparation, we learned of a submitted paper by \cite{kipp19} on this system. We conducted the work here independently. After submitting this manuscript and revising in response to the referee report, we read \citet{kipp19} study in order to compare our results. Our solutions are generally consistent. We infer a larger range of possible masses and eccentricities. We find a smaller radius for the outer planet due to our different stellar parameters derived from ground based spectroscopy and a larger range of possible radii and impact parameters for the inner planet. Ground-based transits aided our work by extending the TTV baseline and filling in transit times that were missed by \TESS.}

\subsection{Summary}
\label{subsec:compare}

 From the TTVs alone, we end up with solutions that occupy two qualitatively different parts of parameter space. The first corresponds to a sub-Saturn{-mass planet} and Neptune{-mass planet} with larger free eccentricities, period ratio near 2.00, and near but not in orbital resonance. The second corresponds to a Jupiter accompanied by a sub-Saturn with smaller free eccentricities, period ratios near 2.02, and librating in orbital resonance. Although the masses are not precisely constrained due to the degeneracy with eccentricity, we narrow the range of possible masses sufficiently to consider these candidates now confirmed as planets.
 
Although we cannot yet rule out the former solution, the latter solution has several appealing features. The period ratio falls outside the observed gap among Kepler multis \citep{fabr14}. The lower free eccentricities and libration of the resonant argument are suggestive of a dissipative process, such as disk migration, capturing the planets into resonance so that we observe them near a 2:1 period ratio. The masses are more typical of the observed radii (Fig.~\ref{fig:rm}).

\section{Discussion}
\label{sec:discuss}

TOI-216 is a system of two known transiting candidates in or near a 2:1 orbital resonance with { accuracy-to-minutes} constraints on their mid-transit times. Unlike most\footnote{See \citealt{daws14} for an example of a \Kepler warm Jupiter with ground-based mid-transit times.} \Kepler systems, the { 12.393 $V$ magnitude star is sufficiently bright} for ground-based follow up to play an important role in supplying additional transits and transit exclusion intervals. From the phases of the TTVs, we identified that the pair contains significant free eccentricity that leads to degeneracy between eccentricities and masses. We ruled out the lowest-mass solutions using the ``photoeccentric'' effect and the highest-mass solutions using transit exclusion intervals from missed ground-based transits. Their mutual inclination may be { modest} (minimum $1^\circ.90^{+0.15}_{-0.34}$) but the component parallel to the sky plane is unknown. We identified two families of solutions. One solution family corresponds to lower masses (a sub-Saturn{-mass planet} and Neptune{-mass planet}), larger eccentricities, period ratio near 2, planets near but not in resonance, and puffy radii. The other corresponds to larger masses (Jupiter{-mass planet} and sub-Saturn{-mass planet}), lower eccentricities, a period ratio of 2.02, masses typical of the planets' sizes, and orbital mean motion resonant libration. We prefer the second family of solutions but cannot yet rule out the first. 

\subsection{Formation and evolution}

TOI-216 joins the population of systems featuring warm, large exoplanets that could not have achieved their close-in orbits through high eccentricity tidal migration (Fig. \ref{fig:arch}). They may have formed at or near their current locations (e.g., \citealt{huan16}), or formed at wider separations and migrated in (e.g., \citealt{lee02}). Both scenarios could lead to planets in or near resonance (e.g., \citealt{dong16,macd18}). The in situ scenario would require the planets to coincidentally form with a period ratio close to 2, but in situ formation sculpted by stability can produce ratios near this value (e.g., \citealt{daws16}). For the lowest-mass solutions, formation beyond the snow line may be necessary to account for the large radii \citep{lee16}.

The planets have at least small and possibly moderate free eccentricities and mutual inclination. The free eccentricities and inclinations might result from dynamical interactions with other undetected planets in the system. For the higher-mass/low eccentricity solution, the eccentricities/inclinations are small enough to be consistent with self-stirring (e.g., \citealt{petr14}) by Neptune-mass or larger planets. The free eccentricities could even be generated by the gas disk (e.g., \citealt{duff15}). However, the free eccentricities in the lower-mass solution would require nearby, undetected giant planets to accompany the observed sub-Saturn{-mass planet} and Neptune{-mass planet} pair. 

Among the eleven systems featuring a warm, large exoplanet with companions with $< 100$~day orbital period (Fig.~\ref{fig:arch}), only TOI-216 and Kepler-30 lack a detected small, short period planet (Section~\ref{subsec:search}). Whatever formation and migration scenario led to the short period planets in the other systems may not have operated here, or the planet may have been lost through stellar collision or tidal disruption. If present but non-transiting, such a planet would need to be mutually inclined to the rest of the system (for example, a non-transiting 3~day TOI-216\,d would need to be inclined by $5^\circ$ with respect to TOI-216\,c). The same stirring environment that led to free eccentricities could also have generated a mutual inclination for this interior planet. (Of course, it may be that no planet formed or migrated interior to TOI-216\,b.) More generally, the mutual inclination between b and c makes it plausible that there are non-transiting planets in the system. 

\subsection{Future observations}

Future \TESS sectors will allow for additional transit timing measurements. As shown in Section~\ref{subsec:predict}, distinguishing between the two families of solutions { may be possible with additional transits of the outer planets. Moreover,} we can likely distinguish between the two families of solutions by measuring the masses through RV follow up: the radial velocity amplitudes are $\sim 53$~m/s and $\sim$~2015~m/s for planet \inn and \out respectively in Solution 1 (Table~\ref{tab:ttv216}) and $\sim$~10~m/s and $\sim$~670~m/s for planet \inn and \out respectively in Solution 2 (Table~\ref{tab:ttv216}). We caution that because of the planets' period ratio and mass ordering, the RV signal alone is subject to significant degeneracy between the inner planet's mass and the outer planet's eccentricity \citep{angl10}.  Combining TTVs and RVs can break this degeneracy.

 { Unfortunately \thisstar does not fall within the observable part of the sky for CHEOPS. Other space-based follow up possibilities, particularly to detect a change in transit depth/impact parameter for the inner planet due to its precession, include Spitzer.}

We expect ground-based observations to play an essential role in follow up of TOI-216. As demonstrated here, ground-based observations can provide accurate and precise transit times for this bright star with two large transiting planets. { For the larger planet in particular, ground-based transits can yield transit times that are more precise than from \TESS data (e.g., the transit observed by LCO in Table \ref{tab:216}). We can identify in advance which transit epoch(s) would be most valuable for distinguishing among models \citep{gold18}. Ground-based transits will allow for a long baseline of observations for better constraining the planets' masses and eccentricities and possibly even detect precession of the planets' orbits.

\acknowledgments
 We thank the \TESS Mission team and follow up working group for the valuable dataset. We acknowledge the use of public \TESS Alert data from pipelines at the \TESS Science Office and at the \TESS Science Processing Operations Center.  This paper includes data collected by the \TESS mission, which are publicly available from the Mikulski Archive for Space Telescopes (MAST). 
 
 This research has made use of the Exoplanet Follow-up Observation Program website, which is operated by the California Institute of Technology, under contract with the National Aeronautics and Space Administration under the Exoplanet Exploration Program. This work has made use of observations from the Las Cumbres Observatory network.  This work has made use of data from the European Space Agency (ESA) mission {\it Gaia} (\url{https://www.cosmos.esa.int/gaia}), processed by the {\it Gaia} Data Processing and Analysis Consortium (DPAC, \url{https://www.cosmos.esa.int/web/gaia/dpac/consortium}). Funding for the DPAC has been provided by national institutions, in particular the institutions participating in the {\it Gaia} Multilateral Agreement. Resources supporting this work were provided by the NASA High-End Computing (HEC) Program through the NASA Advanced Supercomputing (NAS) Division at Ames Research Center for the production of the SPOC data products.
 
 We gratefully acknowledge support by NASA XRP NNX16AB50G and NASA \TESS GO 80NSSC18K1695. The Center for Exoplanets and Habitable Worlds is supported by the Pennsylvania State University, the Eberly College of Science, and the Pennsylvania Space Grant Consortium.  T.D. acknowledges support from MIT's Kavli Institute as a Kavli postdoctoral fellow. K.H. acknowledges support from STFC grant ST/R000824/1. M{\v Z} acknowledges funding from the Australian Research Council (grant DP170102233).
 
 We thank Samuel Hadden and Sarah Morrison for helpful discussions. { We thank the referee for a helpful report that improved the clarity of the paper.}

\bibliographystyle{apj}
\bibliography{biblio}

\end{document}